\documentclass[10pt,a4paper,twocolumn,tightenlines,amsmath,amssymb,nofootinbib,superscriptaddress]{revtex4-2}

\usepackage{graphicx}

\usepackage{amsthm}
\newtheorem*{rep@lemma}{\rep@title}
\newcommand{\newreplemma}[2]{%
\newenvironment{rep#1}[1]{%
 \def\rep@title{#2 \ref{##1}}%
 \begin{rep@lemma}}%
 {\end{rep@lemma}}}
\makeatother

\newreplemma{lemma}{Lemma}

\newtheorem*{definition*}{Definition}
\newtheorem*{theorem*}{Central Result}
\newtheorem*{lemma*}{Lemma}
\newtheorem*{corollary*}{Corollary}

\makeatletter
\DeclareRobustCommand{\cev}[1]{%
  \mathpalette\do@cev{#1}%
}
\newcommand{\do@cev}[2]{%
  \fix@cev{#1}{+}%
  \reflectbox{$\m@th#1\vec{\reflectbox{$\fix@cev{#1}{-}\m@th#1#2\fix@cev{#1}{+}$}}$}%
  \fix@cev{#1}{-}%
}
\newcommand{\fix@cev}[2]{%
  \ifx#1\displaystyle
    \mkern#23mu
  \else
    \ifx#1\textstyle
      \mkern#23mu
    \else
      \ifx#1\scriptstyle
        \mkern#22mu
      \else
        \mkern#22mu
      \fi
    \fi
  \fi
}

\makeatletter
\newcommand*{\balancecolsandclearpage}{%
  \close@column@grid
  \clearpage
  \twocolumngrid
}
\makeatother

\newcommand{\ket}[1]{\left | #1 \right\rangle}
\newcommand{\bra}[1]{\left \langle #1 \right |}

\newcommand{\braket}[2]{\left\langle #1|#2\right\rangle}
\newcommand{\ketbra}[2]{|#1\left\rangle\right\langle #2 |}

\newcommand{\past}[1]{\cev{#1}}

\newcommand{\inlineheading}[1]{\textbf{{#1}}}

\usepackage{color}
\usepackage[dvipsnames]{xcolor}
\usepackage[normalem]{ulem}
\definecolor{blue}{rgb}{0,0.2,1}

\definecolor{red}{rgb}{0.9,0,0}

\definecolor{green}{rgb}{0,0.7,0}

\newcommand{\CQT}{Centre for Quantum~Technologies, National~University~of~Singapore, 3 Science Drive 2, Singapore 117543}
\newcommand{\Oxf}{Department of Physics, Clarendon~Laboratory, University of Oxford, Parks Road, Oxford, OX1 3PU, United Kingdom.}
\newcommand{\NTU}{School of Physical and Mathematical Sciences, Nanyang Technological University, Singapore 637371}
\newcommand{\complexity}{Complexity Institute, Nanyang Technological University, Singapore 637335}
\newcommand{\IQOQI}{Institute for Quantum Optics and Quantum Information, Austrian Academy of Sciences, Boltzmanngasse 3, A-1090 Vienna, Austria}
\newcommand{\ucdavis}{Complexity Sciences Center and Physics Department, University of California at Davis, One Shields Avenue, Davis, California 95616, USA}
\newcommand{\SMU}{School of Computing and Information Systems, Singapore Management University, 80 Stamford Road, Singapore 178902}

\begin{document}

\title{Surveying structural complexity in quantum many-body systems}

\author{Whei Yeap Suen}
\email{wheiyeap@u.nus.edu}
\affiliation{\SMU}
\affiliation{\CQT}

\author{Thomas J.~Elliott}
\email{physics@tjelliott.net}
\affiliation{Department of Mathematics, Imperial College London, London SW7 2AZ, United Kingdom}
\affiliation{\complexity}
\affiliation{\NTU}

\author{Jayne Thompson}
\affiliation{\CQT}

\author{\mbox{Andrew J.~P.~Garner}}
\affiliation{\IQOQI}
\affiliation{\NTU}
\affiliation{\CQT}

\author{John R. Mahoney}
\affiliation{\ucdavis}

\author{Vlatko Vedral}
\affiliation{\Oxf}
\affiliation{\CQT}
\affiliation{Department of Physics, National University of Singapore, 3 Science Drive 2, Singapore 117543}

\author{Mile Gu}
\email{cqtmileg@nus.edu.sg}
\affiliation{\NTU}
\affiliation{\complexity}
\affiliation{\CQT}

\date{\today}

\begin{abstract}
Quantum many-body systems exhibit a rich and diverse range of exotic behaviours, owing to their underlying non-classical structure. These systems present a deep structure beyond those that can be captured by measures of correlation and entanglement alone. Using tools from complexity science, we characterise such structure. We investigate the structural complexities that can be found within the patterns that manifest from the observational data of these systems. In particular, using two prototypical quantum many-body systems as test cases -- the one-dimensional quantum Ising and Bose-Hubbard models -- we explore how different information-theoretic measures of complexity are able to identify different features of such patterns. This work furthers the understanding of fully-quantum notions of structure and complexity in quantum systems and dynamics.
\end{abstract}

\maketitle

\section{Introduction}
Quantum many-body systems hold a distinguished position in modern physics, playing a vital role in providing insight into the physical world. On the one hand, they constitute an excellent platform for studying a range of phenomena through their utility in quantum simulation~\cite{feynman1982simulating, bloch2012quantum, lewenstein2012ultracold, johnson2014what}. Conversely, the properties intrinsic to these systems are interesting in their own right, and they thus form a target of simulation and modelling themselves~\cite{jaksch1998cold, greiner2002quantum, jaksch2003creation, bloch2008many, baumann2010dicke, ritsch2013cold, aidelsburger2013realization, miyake2013realizing}. Understanding the structures present in these systems, and the resources needed to characterise, study and emulate them, is thus of paramount importance.

Quantum entanglement~\cite{amico2008entanglement, horodecki2009quantum} captures the quantum correlations present in a system, and so plays a significant role in identifying structure in quantum systems. In particular, the \emph{half-chain entanglement} quantifies the amount of information shared between the left and right halves of a one-dimensional quantum system; it provides an indicator to the amount of classical resources needed to simulate such systems~\cite{vidal2003efficient, verstraete2006matrix, schollwock2011density, orus2014practical}, and related quantum mutual information-based quantities have previously been associated with structural complexity~\cite{valdez2017quantifying}. Nevertheless, complex systems possess structure beyond such correlations; we turn to tools from complexity science to identify and quantify this structure. The field of \emph{computational mechanics}~\cite{crutchfield1989inferring, shalizi2001computational, crutchfield2012between} adopts an information-theoretic approach to this this task, and equates structure in a stochastic process with the minimal amount of information that must be stored by a model that replicates its behaviour. Moreover, it offers a systematic approach to determining such minimal models.

Impelled by the growth of quantum technologies, recent efforts have extended the computational mechanics framework into the quantum regime, finding that classical limits on the information that models must store can be overcome~\cite{gu2012quantum, mahoney2016occam, palsson2017experimentally, aghamohammadi2018extreme, binder2018practical, elliott2018superior, elliott2019memory, liu2019optimal}. This fundamentally changes how we might perceive and characterise structure -- two prominent examples being the ambiguities of simplicity~\cite{suen2017classical,aghamohammadi2017ambiguity,jouneghani2017observing} and optimality~\cite{loomis2019strong, liu2019optimal}, which highlight properties of complexity that might be considered truisms classically no longer hold in the quantum domain. Practically, these quantum models can provide memory savings for stochastic simulation, and the resource gap between minimal quantum and classical simulators can even grow unbounded~\cite{garner2017provably, aghamohammadi2017extreme, elliott2018superior, elliott2019memory, thompson2018causal, elliott2020extreme, elliott2021quantum}.

It is natural then to ask how these measures of complexity look -- and what they can tell us about structure -- when applied to quantum many-body systems. In this article, we apply this framework to study their structure and complexity. To this end, we look at the structure manifest in the measurement statistics of quantum states. This serves as a crucial first step in identifying the structure in the quantum processes that gave rise to the states. We begin by reviewing in Section \ref{sec:framework} the relevant details of causal models and measures of complexity that form the background to our work. In Section \ref{section:results}, we introduce the mapping from states of quantum chains to stochastic processes through the statistics of observation sequences, and apply it to quantify structure in the one-dimensional quantum Ising and Bose-Hubbard models. We discuss the implications of our results and the future directions to which our framework may be applied in Section \ref{sec:discussion}.

\section{Framework}\label{sec:framework}
\textbf{Stochastic processes.} We consider discrete-event, discrete-time stationary stochastic processes~\cite{khintchine1934korrelationstheorie}. At each timestep $t\in \mathbb{Z}$ such a process emits a symbol $r_t$ drawn from a configuration space $\mathcal{R}$. We use $R_t$ to denote the random variable governing output $r_t$. We also designate the semi-infinite strings $\cev{R}_t := \dots R_{t-1}R_t$ and $\vec{R} _t:= R_{t+1}R_{t+2}\dots$ as the random variables associated with the past and future observation sequences $\cev{r}_t := \dots r_{t-1}r_t$ and $\vec{r} _t:= r_{t+1}r_{t+2}\dots$ at time $t$ respectively (throughout, upper case indicates random variables, and lower case the corresponding variates). The output symbol statistics are then described by a conditional probability distribution over these strings $P(\vec{R}_t | \cev{R}_t)$, detailing how future observations are correlated with past observations. We use the notation $r_{t:t+n}=r_tr_{t+1}\dots r_{t+n-1}$ to represent the sequence of outputs between $t$ and $t+n-1$. Stationarity of a stochastic process is defined by $P(R_{0:n})=P(R_{L:L+n})$ $\forall n,L \in \mathbb{Z}$; this allows us to drop the subscript $t$ from semi-infinite strings.

\textbf{Causal models.}  A \emph{causal model} of a stationary stochastic process~\cite{crutchfield1989inferring, shalizi2001computational, thompson2018causal} is tasked with replicating its future output statistics according to the distribution $P(\vec{R} | \cev{R})$; it stores information about the past of a stochastic process in its internal memory, and uses this to predict the future output statistics. Crucially, a causal model contains no information about the future that cannot be inferred from the past (i.e., the mutual information between the memory and future outputs given the past outputs is zero). Causal models use an encoding function $f$ to map pasts $\cev{r}$ to states $m\in\mathcal{M}$ according to $m=f(\cev{r})$, such that $P(\vec{R}|m=f(\cev{r}))=P(\vec{R}|\cev{r})$. At each timestep $t$, the model produces an output $r$ following $P(R_{t+1}=r|m_t)=P(R_{t+1}=r|\cev{r}_t)$. At time $t+1$, $r$ becomes part of the past observation sequence, and the memory state is updated to $m_{t+1}=f(\cev{r} ')$, where the new past $\cev{r}'=\cev{r} r$ is the concatenation of the previous past with the new output symbol. The information stored by such a model is given by the Shannon entropy of its set of internal memory states: 
\begin{equation}
H(M)=\sum_{m\in \mathcal{M}}{-P_m\log{P_m}},\label{memory_c}
\end{equation}
where $P_m=\sum_{\cev{r}\in m} P(\cev{r})$.

For any given stationary stochastic process, one can construct myriad causal models of the process that will faithfully replicate the future statistics. The field of computational mechanics provides us with a systematic way to identify and construct the provably optimal classical causal model of a given process~\cite{crutchfield1989inferring, shalizi2001computational, crutchfield2012between}. By optimal, we here mean the model that stores the least possible amount of past information [Eq.~\eqref{memory_c}] while accurately simulating the process; the optimal classical causal model is referred to as the $\varepsilon$-machine. In this framework, sets of pasts are grouped into equivalence classes called \textit{causal states} according to the relation
\begin{align}
\cev{r} \sim_e \cev{r}' \iff P(\vec{R}=\vec{r}|\cev{R}=\cev{r})=P(\vec{R}=\vec{r}|\cev{R}=\cev{r}') \forall \vec{r}. \label{eq_eqn}
\end{align}
Eq.~\eqref{eq_eqn} mandates that past observations leading to statistically identical futures belong to the same causal state. Let us denote $\mathcal{S}$ as the set of causal states, where each state $s\in\mathcal{S}$ is given by an associated encoding function $s =\epsilon(\past{r})$. At each time step the $\varepsilon$-machine transitions from causal state $j$ to $k$ whilst emitting an output $r\in\mathcal{R}$, with transition probability $T_{kj}^r=P(R_{t+1}=r,S_{t+1}=k|S_t=j)$. The encoding function enforces \textit{unifilarity} of the $\varepsilon$-machine, where, given the current causal state $j$ and the emitted output symbol $r$, the subsequent causal state $k$ of the model is uniquely specified~\cite{shalizi2001computational}. We denote this mapping by a function $\lambda(j,r)$ that outputs the value of the label of the subsequent causal state. This allows us to express
\begin{align}
T_{kj}^r &= P(R_{t+1}=r,S_{t+1}=k|S_t=j)\\
&= P\left( r| j \right)\delta_{k,\lambda(j,r)} \label{plambda}
\end{align}
where $\delta_{jk}$ is the Kronecker-$\delta$ function. The amount of information required to track the dynamics of causal states has been widely employed as a measure of structure~\cite{crutchfield1989inferring, crutchfield1997statistical, palmer2000complexity, varn2002discovering, clarke2003application, park2007complexity, li2008multiscale, haslinger2010computational, kelly2012new, munoz2020general}, designated as the \textit{statistical complexity}:
\begin{equation}
C_\mu:=-\sum_{j}P_j\log{P_j}, \label{cm}
\end{equation}
where $P_j=\sum_{\cev{r}\in j} P(\cev{r}) $ is the steady-state probability of causal state $j$. The statistical complexity is often compared to the mutual information between the past and future of the system,
\begin{equation}
E = \sum_{\cev{r},\vec{r}} P(\cev{r}, \vec{r}) \log \left( \frac{P(\cev{r},\vec{r})}{P(\cev{r})P(\vec{r})} \right).
\end{equation}
a quantity known as the \textit{excess entropy}. It quantifies the amount of information in the past that correlates with future statistics. The excess entropy (also sometimes called the `predictive information') is also used as a measure of complexity~\cite{shaw1984dripping, bialek2001predictability}. The data processing inequality~\cite{nielsen2010quantum} ensures that the excess entropy represents a lower bound on the amount of information a simulator of a process must store in any physical theory, and thus $C_\mu\geq E$.

\textbf{Quantum causal models.} Recently, computational mechanics has been extended into the quantum regime~\cite{gu2012quantum}, where it has been shown that quantum effects allow one to construct causal models that require a lower amount of information than is classically possible. The present state-of-the-art systematic construction methods for quantum models~\cite{aghamohammadi2018extreme, binder2018practical, liu2019optimal} involve step-wise unitary interactions between the model memory and a probe system. The memory of such quantum models store a member of a set of quantum memory states $\{ \ket{s_j} \}$ that are in one-to-one correspondence with the causal states of the process. The memory states are then used to produce the future outputs of the process sequentially. Starting from state $\ket{s_j}$ and a blank ancilla, there exists a unitary operator $U$ that satisfies

\begin{align}
U\ket{s_j}\ket{0} &= \sum_{r\in\mathcal{R}}\sqrt{P(r|j)}\ket{s_{\lambda(j,r)}}\ket{r}. \label{unitary}
\end{align}
The stationary state of the model memory is given by $\rho=\sum_j{P_j \ket{s_j}\bra{s_j}}$, where $P_j$ is as defined for Eq.~\eqref{cm}. The amount of information stored internally is given by the von Neumann entropy of $\rho$:
\begin{equation}
C_q := -\text{Tr}\left( \rho \log_2 {\rho} \right). \label{cq}
\end{equation}
$C_q$ is referred to as the \emph{quantum statistical memory}. In general, the memory states are non-orthogonal (i.e., $\braket{s_j}{s_k}\geq 0$), due to a quantum model being able to encode pasts with partially overlapping futures into partially overlapping states. Therefore, $C_q \leq C_\mu$;  a quantum causal model can utilise less memory than the $\varepsilon$-machine of the same process~\cite{gu2012quantum, mahoney2016occam, palsson2017experimentally, aghamohammadi2018extreme, binder2018practical, elliott2018superior, elliott2019memory, liu2019optimal}. While the quantum models presented here require less memory than $\varepsilon$-machines, they are in general not optimal over all quantum models~\cite{liu2019optimal}. However, in certain specific cases, this construction has been shown to be the provably optimal quantum model~\cite{suen2017classical,thompson2018causal}. Paralleling $C_\mu$, the minimal possible memory cost across all quantum models is called the quantum statistical complexity. $C_q$ thus upper bounds the quantum statistical complexity, and due to the difficulty associated with finding the true quantum minimum has been suggested as a potential measure of complexity in itself~\cite{gu2012quantum, tan2014towards, ho2020robust, ho2021quantum}. As with $C_\mu$, the data processing inequality mandates that $C_q\geq E$.

\textbf{Measures of complexity.} We focus our attention on the following measures of complexity:
\begin{itemize}
\item \textit{Statistical complexity} $C_\mu$ 
\item \textit{Quantum statistical memory} $C_q$
\item \textit{Excess entropy} $E$
\item \textit{Half-chain entanglement entropy} $S_{\frac{1}{2}}$
\end{itemize} 
$C_\mu$, $C_q$, and $E$ have been formally introduced above. The half-chain entanglement entropy of a one-dimensional quantum many-body system quantifies the amount of quantum correlations (entanglement) between the left and right halves of the system~\cite{amico2008entanglement}. It is defined as
\begin{align}
S_{\frac{1}{2}}& :=-\text{Tr}(\rho_A \log_2 \rho_A) =-\text{Tr}(\rho_B \log_2 \rho_B),
\end{align}
where $\rho_A=\text{Tr}_B\left( \rho_{AB} \right)$ and $\rho_B=\text{Tr}_A\left(\rho_{AB} \right)$ are the density matrices describing the states of the left and right halves of the quantum system respectively. $S_{\frac{1}{2}}$ is measurement basis-independent, and depends only on the state of the quantum many-body system. $S_\frac{1}{2}$ is very closely related to the quantum mutual information ($I_q\left( A, B \right)=2S_{\frac{1}{2}}$ for pure states), and is loosely analogous to the excess entropy for quantum processes. Beyond identifying correlations, entanglement has also been suggested as an indicator of critical points of phase transitions in quantum many-body systems~\cite{osterloh2002scaling, osborne2002entanglement, vidal2003entanglement}. However, experimental measurement of $S_\frac{1}{2}$ in real quantum systems is a highly non-trivial task~\cite{daley2012measuring, abanin2012measuring, islam2015measuring, elliott2015multipartite}.

\section{Results}\label{section:results}
\textbf{Stochastic processes from quantum many-body systems.}
The measures of structural complexity used in this manuscript are defined in terms of classical stochastic processes. 
To apply them to quantum systems, we need some method to meaningfully extract such a process from a quantum system.
Since measurements of a quantum system are inherently probabilistic, when making a series of measurements on a quantum system the outputs form a stochastic sequence. We can then analyse the structure in this sequence using the above measures. The structure in this sequence can embody structure present in the underlying quantum state that gave rise to the observations. Taking this as a proxy for structure in the quantum system itself, the approach can be described as a `semi-classical' method for describing the structural complexity of quantum systems and processes.

\begin{figure}[htb]
	\centering
	\includegraphics[width=0.5\textwidth]{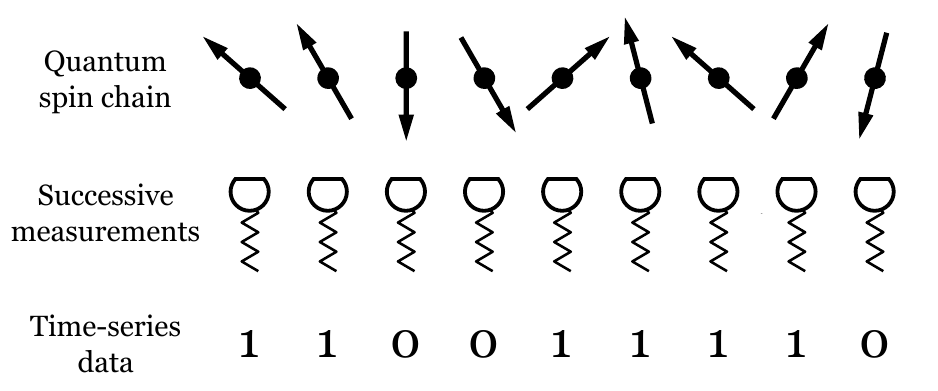}
	\caption{Extracting stochastic processes from quantum chains: By measuring a quantum chain with local operators sequentially at each site a time-series is generated, with temporal indices in the series corresponding directly to spatial indices of the sites. The structure in these sequences can be used as a proxy of structure in the underlying quantum chain.}
	\label{figmeasure}
\end{figure}

There are many possible measurement protocols that could extract a classical time-series from an infinite length one-dimensional quantum many-body chain. 
For instance, one could measure the same site at multiple points in time, allowing the system to evolve between measurements.
Alternatively, as done here, one can take measurements sequentially across consecutive sites of the chain, sweeping from left to right.
This effectively measures one different site per timestep, and the outcomes of this form a time-series, where the temporal indices of the time-series are in one-to-one correspondence with the spatial indices of the sites in the chain. 
This is illustrated in Fig.~\ref{figmeasure}.
 
Specifically, we consider non-degenerate, site-local measurement operators such that on site $j$, a set of measurement outcomes $r_j\in \mathcal{R}$ are associated with unique eigenstates $\ket{r_j}$ of the measurement operator. This measurement outcome is taken to be the corresponding $j$-th observation in the constructed stochastic process. As we sweep the output sequence from left to right, it follows then that the output sequence $\overleftrightarrow{r}$ occurs with probability 
\begin{equation}
P(\overleftrightarrow{r})=\bra{\psi}\bigotimes_{j=-\infty}^{\infty}\ket{r_j}\braket{r_j}{\psi} \label{probability}
\end{equation}
where $\ket{\psi}$ is the quantum state of the one-dimensional system being investigated. 
For the examples considered here, we will take the quantum chains to be in the ground states of their respective Hamiltonians, with no temporal evolution between measurements. 
We emphasise that there is no dynamical evolution of the systems considered here; the temporal dynamics of the extracted stochastic process manifest as a mapping of spatial position in the underlying chain.

\textbf{Numerical considerations.}
For numerical tractability, we make two approximations regarding the system size and the measures of complexity. Firstly, rather than an infinite chain, we study large finite-size quantum chains of length $N$. Secondly, we introduce the truncated Markov memory order $L$. It represents the number of sites from which past information may be obtained. That is, we approximate $P(r|\cev{r})\approx P(r|r_{-L:1})$. We take $N\gg L$. We employ tensor network methods [19] (see Appendix) to obtain near exact ground states of the example systems we study, as well as to extract the corresponding measurement sequences that form the stochastic processes.

We now apply our framework to explore the structures of two paradigmatic quantum many-body chains in their respective ground states.

\textbf{Quantum Ising chain}. A quantum Ising chain~\cite{mattis2006theory} describes the physics of a system of interacting quantum spins subject to the influence of a magnetic field. They are governed by the Hamiltonian:
\begin{equation}
\mathcal{H}_{QI}=\sum_{l}{-J\sigma_l^x \sigma_{l+1}^x - B \sigma_l^z} \label{hamiltonianQI}
\end{equation}
where $J$ is a coupling parameter, $B$ is the external magnetic field strength, and $\sigma_l^w$, $w\in\{x,y,z\}$, are the Pauli operators at site $l$. The system undergoes a quantum phase transition at $B/J=0.5$. At $B \gg J$, the field along the $z$-direction dominates the correlations in the system; the ground state of the system is $\ket{\psi_g} = \ket{\dots \downarrow_{i-1}\downarrow_{i}\downarrow_{i+1} \dots}$, fully-polarised along the $z$-axis.  On the other hand, when $B \ll J$, the field is much weaker than the spin-spin correlation along the $x$-direction. There are then two degenerate ground states, with all spins either parallel or anti-parallel with the $x$-axis, $\ket{\psi_g}=\ket{\dots\rightarrow_{i-1}\rightarrow_{i}\rightarrow_{i+1}\dots}$ or $\ket{\dots\leftarrow_{i-1}\leftarrow_{i}\leftarrow_{i+1}\dots}$. 

We investigate the structure of the model at a range of truncated Markov memory orders $L\in\{ 1, 3, 5, 7, 9 \}$, with $N=500$. The causal states are then determined by the $L$ rightmost spins of the past spins, as defined in Eq.~\eqref{eq_eqn}. We find that each length $L$ spin configuration belongs to a unique causal state. To ensure that there is no spurious splitting of causal states, we determine that the conditional probability distributions differ by more than $\mathcal{O}\left( 10^{-12}\right)$ from each other, while the ground states $\ket{\psi_g}$ are accurate to $\mathcal{O}\left(10^{-14}\right)$; i.e., the error in the ground state is significantly smaller than the distance between conditional distributions of the spin configurations.

Using the framework above, we study the structural complexity of the quantum Ising chain through its measurement sequences as obtained from $\sigma_\theta$-basis measurements, where
\begin{equation}
\sigma_\theta =
\begin{pmatrix}
\cos{\theta} & \sin{\theta} \\
\sin{\theta} & -\cos{\theta}
\end{pmatrix},
\end{equation}
with $\theta \in [0,{\pi}/{2}]$ the angle measured from the $z$-axis of the Bloch sphere. Angles $\theta=0$ and $\theta=\pi/2$ correspond to the $z$- and $x$-axes respectively. Intuitively, $\sigma_z$ measurement in the $B \gg J$ regime will result in a highly-ordered stochastic process, while $\sigma_x$ will return a near-random process. Conversely in the $B \ll J$ regime, measurement sequences along $\sigma_z$ will yield a near-random stochastic process, while $\sigma_x$ will result in a highly-ordered process.

\begin{figure}
	\centering
	\includegraphics[width=0.5\textwidth]{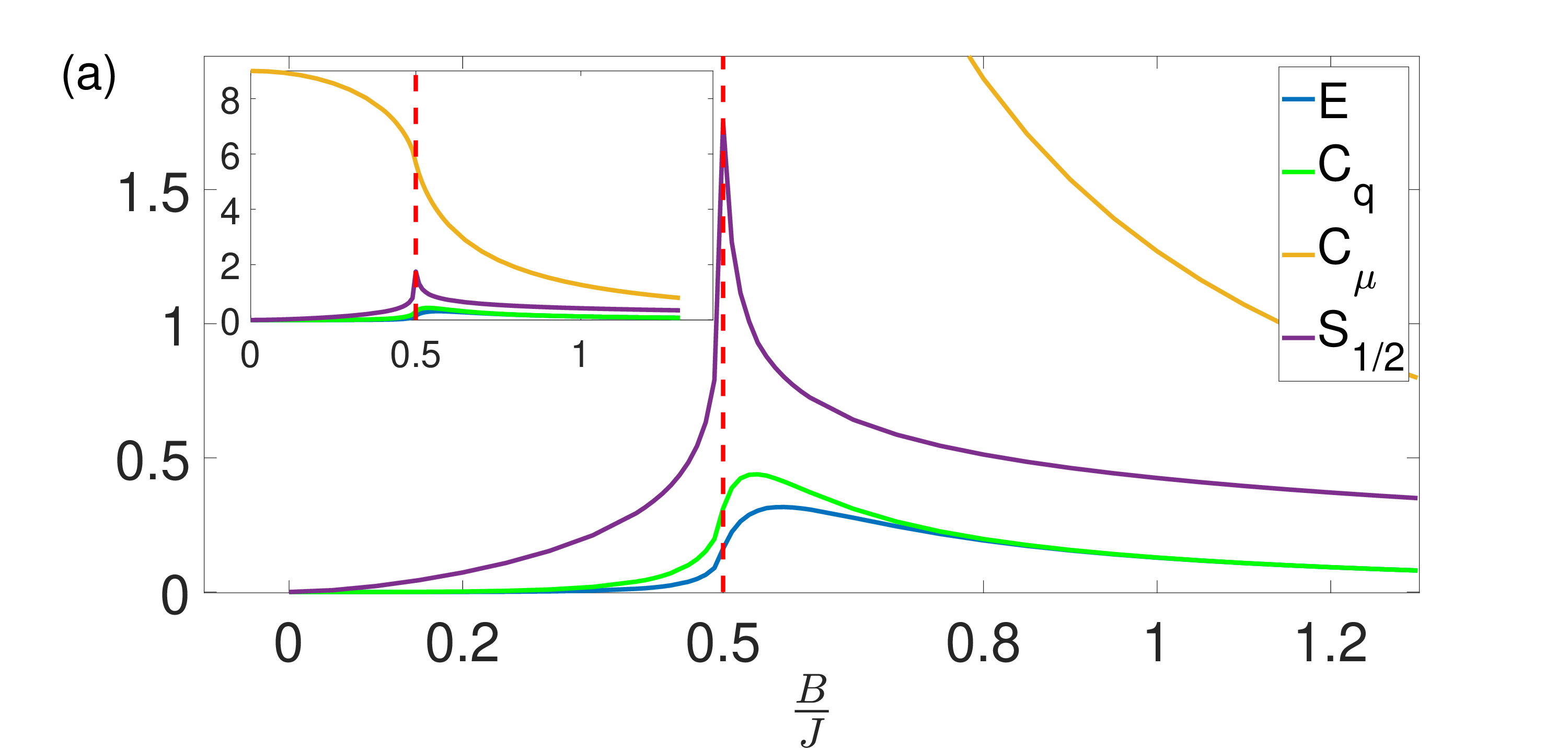}\\
		\includegraphics[width=0.5\textwidth]{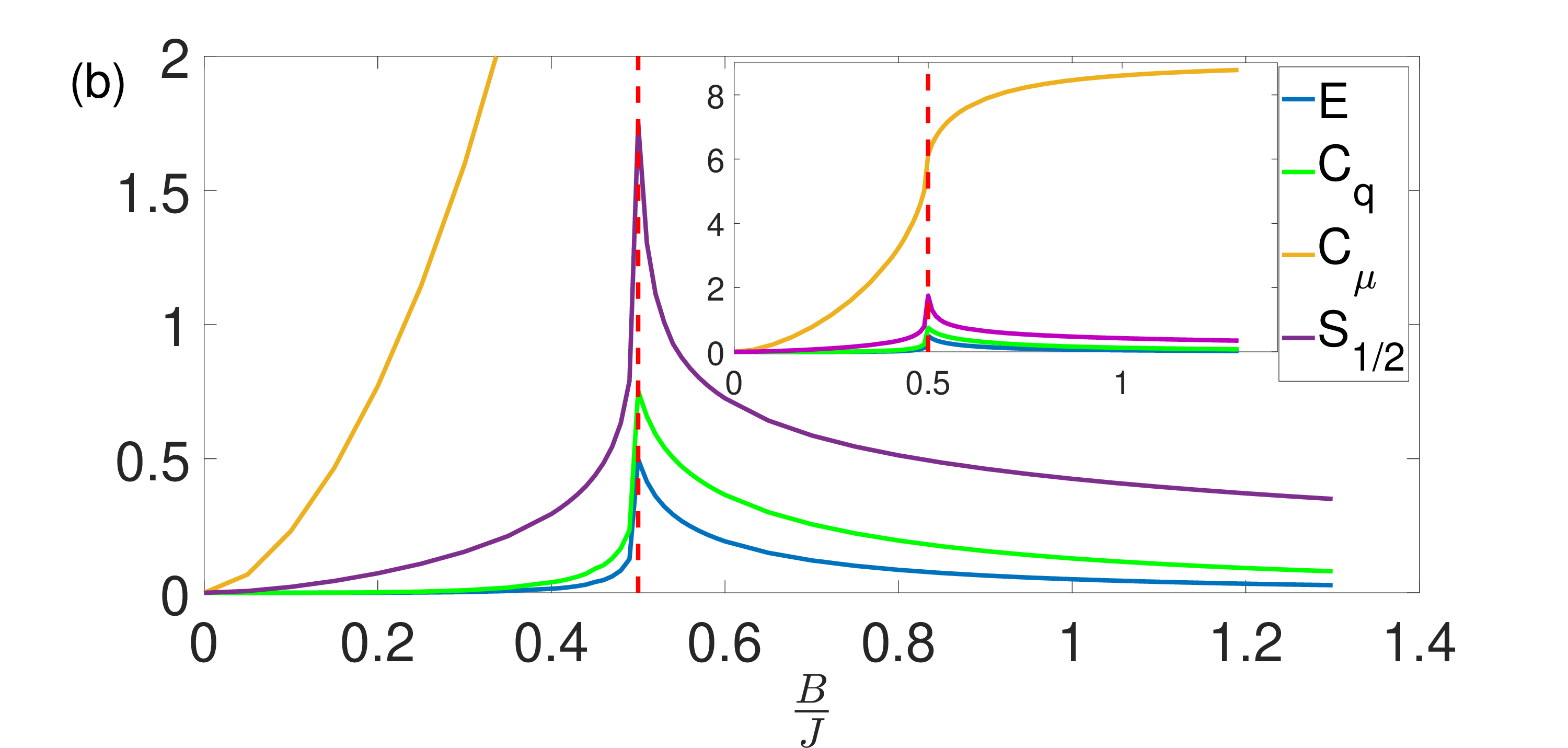}\\
	\caption{Comparison between $E$, $C_q$, $C_\mu$ in the (a) $\sigma_{z}$, and (b) $\sigma_{x}$ measurement bases, and $S_{\frac{1}{2}}$ of the one-dimensional quantum Ising system for $L=9$. The vertical dashed line demarcates the phase transition of the system.}
	\label{fig:isingcompare}
\end{figure}

Fig.~\ref{fig:isingcompare} compares $C_\mu$, $C_q$, $E$, and $S_{\frac{1}{2}}$ for measurements of $\sigma_x$ and $\sigma_z$ at different $B/J$, where we observe interesting differences between these measures of structure and complexity. Firstly, in all measurement bases studied, $C_q$, $E$, and $S_{\frac{1}{2}}$ reach maximal values close to the phase transition $B/J\approx 0.5$. Instead, $C_\mu$ exhibits its largest gradient near the phase transition. 

We also observe that $S_{\frac{1}{2}}> C_q$ for all measurement bases. This is because projecting a  quantum state onto a specific basis effectively destroys information about other measurement bases, while $S_{\frac{1}{2}}$ is a quantity that takes into account the full information contained in the quantum state. The relation $S_{\frac{1}{2}}> C_q$ highlights that simulating a quantum system measured in a specific basis can require less information than the quantum correlations present in the system. This highlights that if we don't require replication of all measurement bases, the state is not the most efficient simulator of itself.

Fig.~\ref{fig:isingcompare} also shows that when sequences from measurement outcomes of the quantum Ising chain are more ordered ($B\gg J$ for $\sigma_{z}$-basis, $B\ll J$ for $\sigma_{x}$-basis), the corresponding values of $C_\mu$, $C_q$, and $E$ are lower. In a highly-ordered stochastic process, the corresponding $\varepsilon$-machine consists of a single dominant causal state that is occupied with very high probability, and other causal states arise with very low probabilities. Thus the resulting $\varepsilon$-machine requires little information to be stored to accurately simulate the corresponding stochastic process. Our quantum model behaves similarly in this regime, hence $C_q$ mirrors $C_\mu$. Further, in this parameter regime the more ordered the sequences are, the less information is carried forward from the left half to the right half of the system, resulting in low values of $E$.

On the other hand, when the observation sequences are near-random ($B\gg J$ for $\sigma_{x}$-basis, $B\ll J$ for $\sigma_{z}$-basis), $C_\mu$ exhibits drastically different qualitative behaviour compared to $C_q$ and $E$, as seen in Fig.~\ref{fig:isingcompare}. Both $C_q$ and $E$ are lower when the sequences appear more random, unlike $C_\mu$, which saturates in this regime. This is because in the near-random limit the past configurations have different-yet-strongly-overlapping conditional future probability distributions, and thus they are mapped into different causal states. As a classical model, the $\varepsilon$-machine can only store information in distinguishable states, despite multiple causal states having significantly overlapping conditional future statistics; consequently, $C_\mu$ is high in this regime. In contrast, quantum models have the ability to store information in non-orthogonal states. Thus, causal states with highly-overlapping future conditional probability distributions will be encoded into highly-overlapping quantum states, resulting in low $C_q$ in the near-random regime.

\begin{figure}
	\centering
	\includegraphics[width=0.5\textwidth]{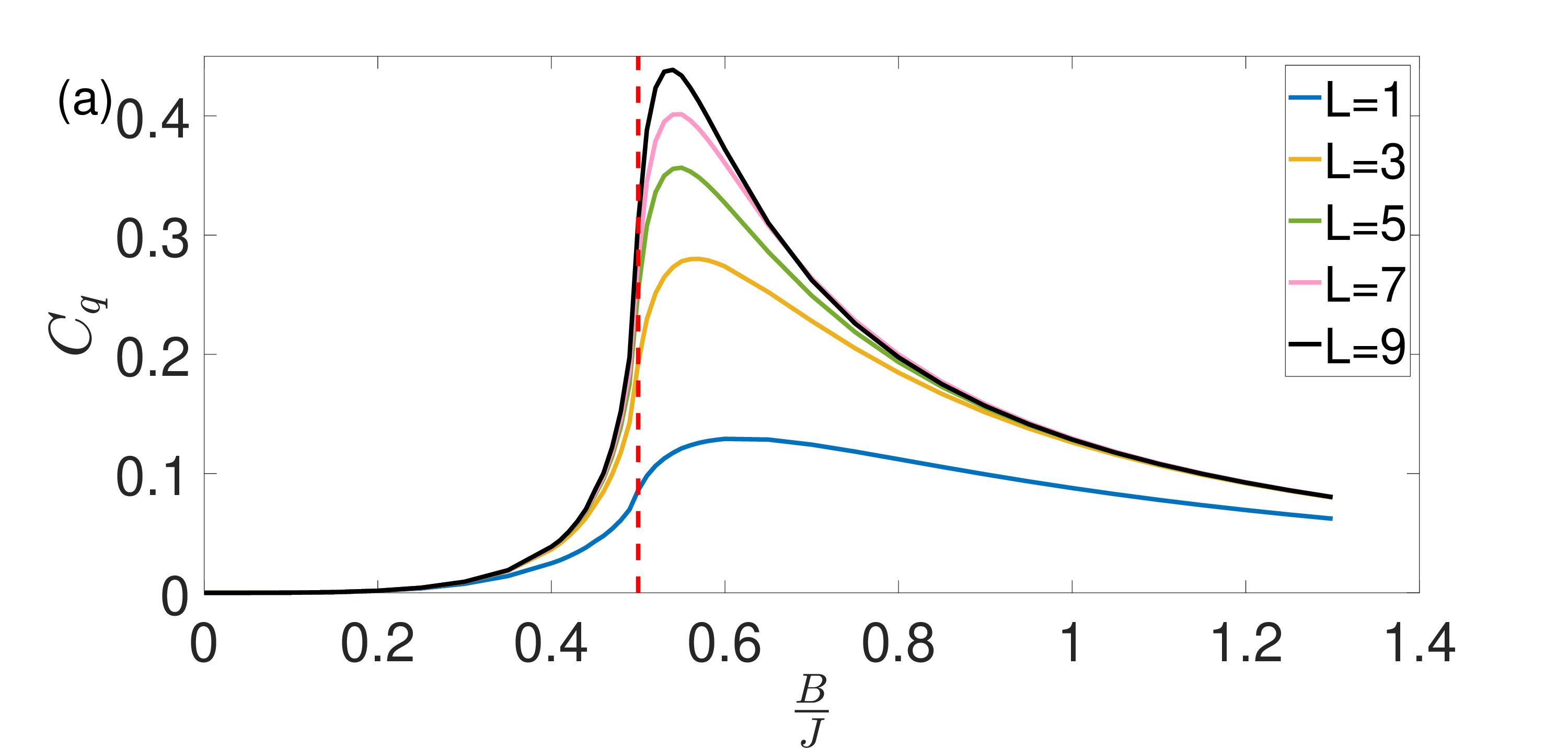}\\
	\includegraphics[width=0.5\textwidth]{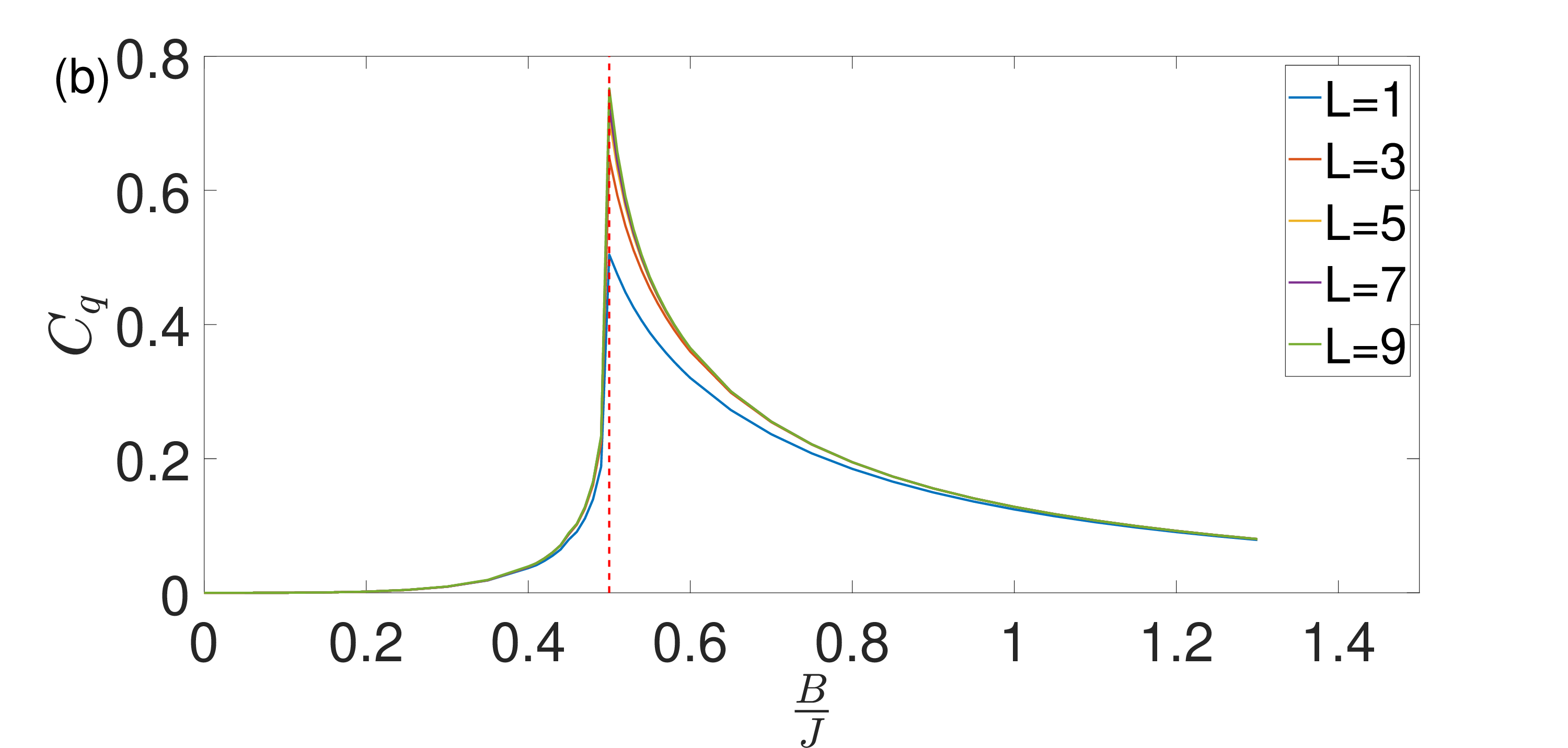}\\
	\caption{$C_q$ plotted against $B/J$ for sequences of measurements of the quantum Ising chain along (a) $\sigma_{z}$, (b) $\sigma_{x}$, . The vertical dashed lines demarcate the phase transition of the system.}
	\label{fig:cq_qi}
\end{figure}

\begin{figure}
	\centering
		\includegraphics[width=0.5\textwidth]{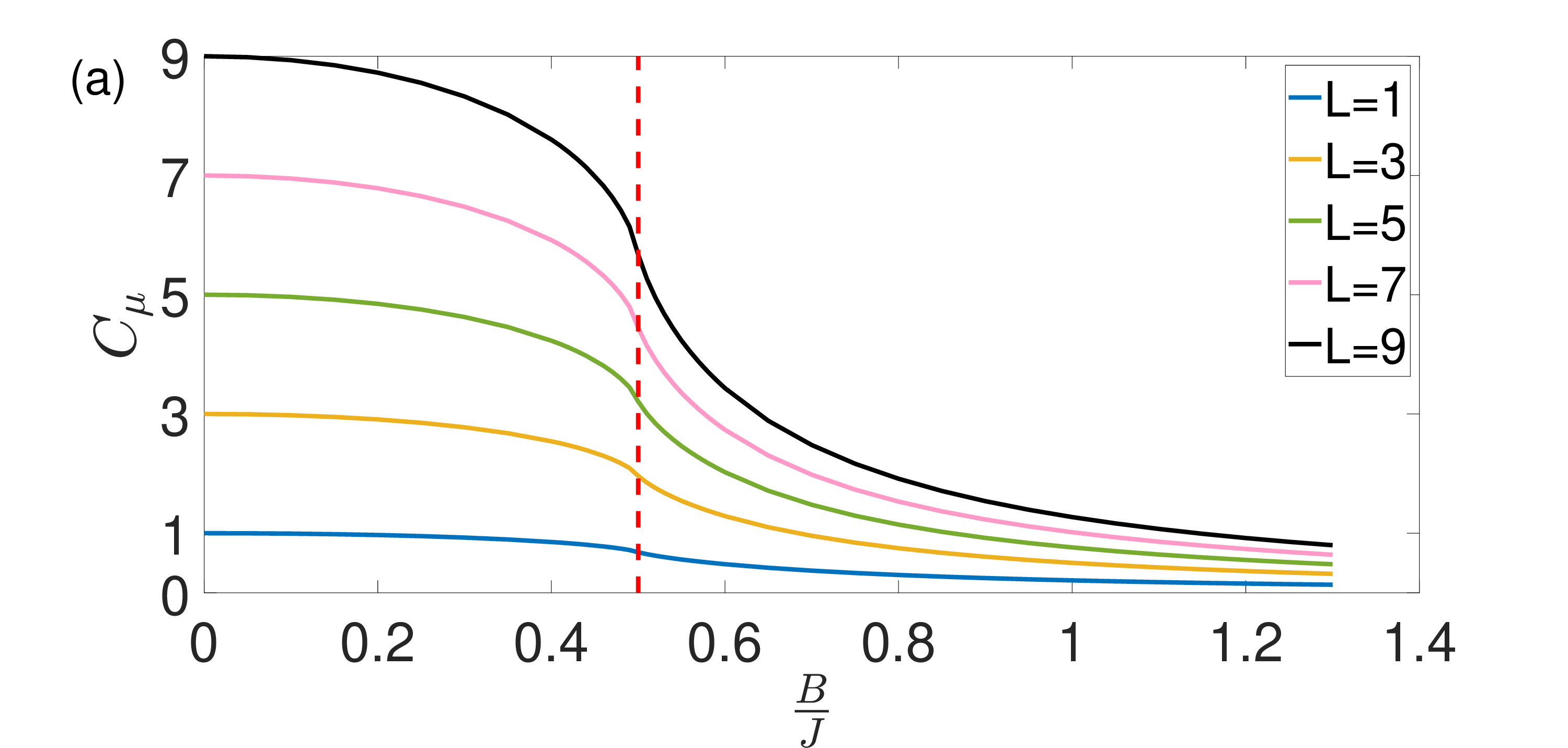}\\
		\includegraphics[width=0.5\textwidth]{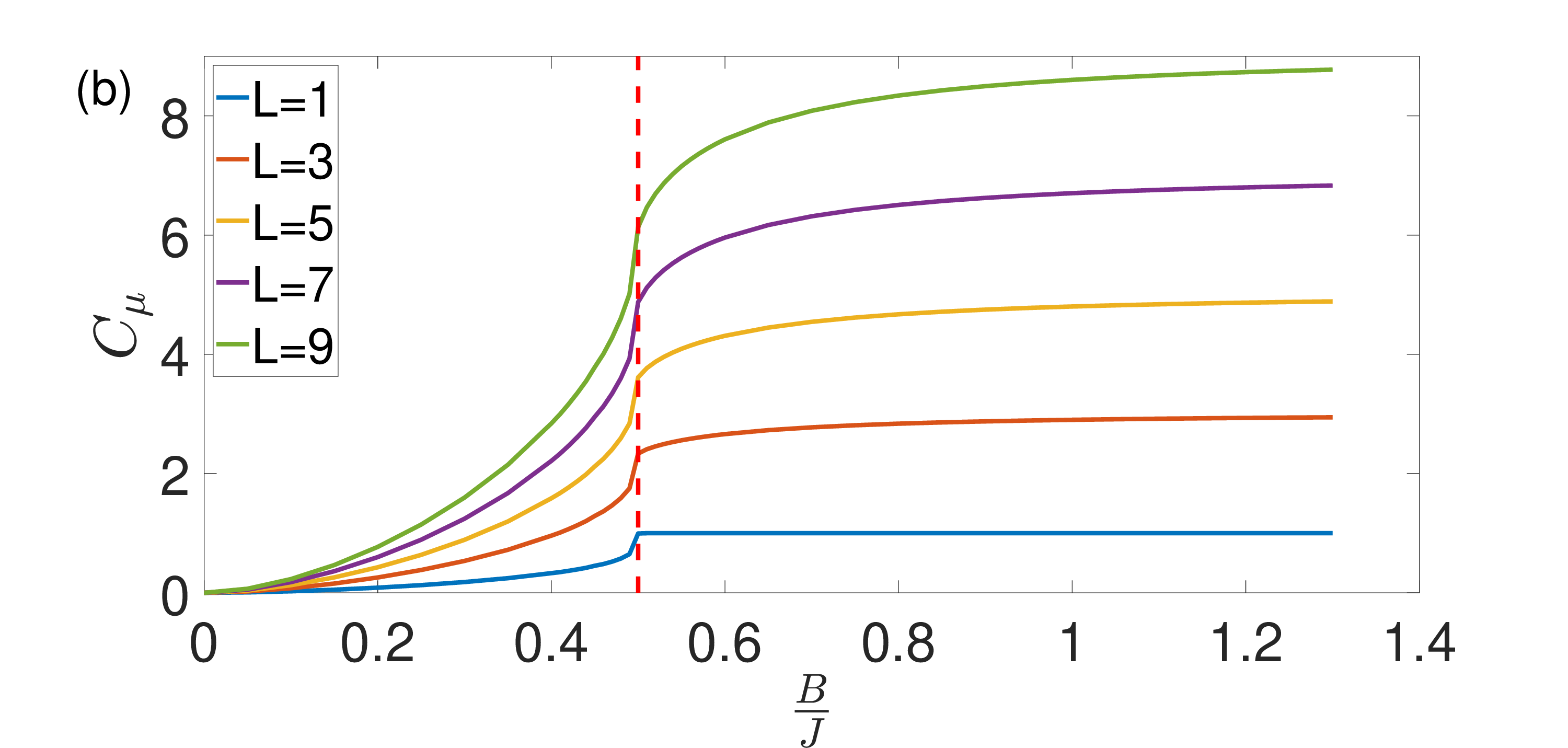}\\
	\caption{$C_\mu$ plotted against $B/J$ for sequences of measurements of the quantum Ising chain along (a) $\sigma_{z}$, (b) $\sigma_{x}$. The vertical dashed lines demarcate the phase transition of the system.}
	\label{fig:cm_qi}
\end{figure}

\begin{figure}
	\centering
	\includegraphics[width=0.5\textwidth]{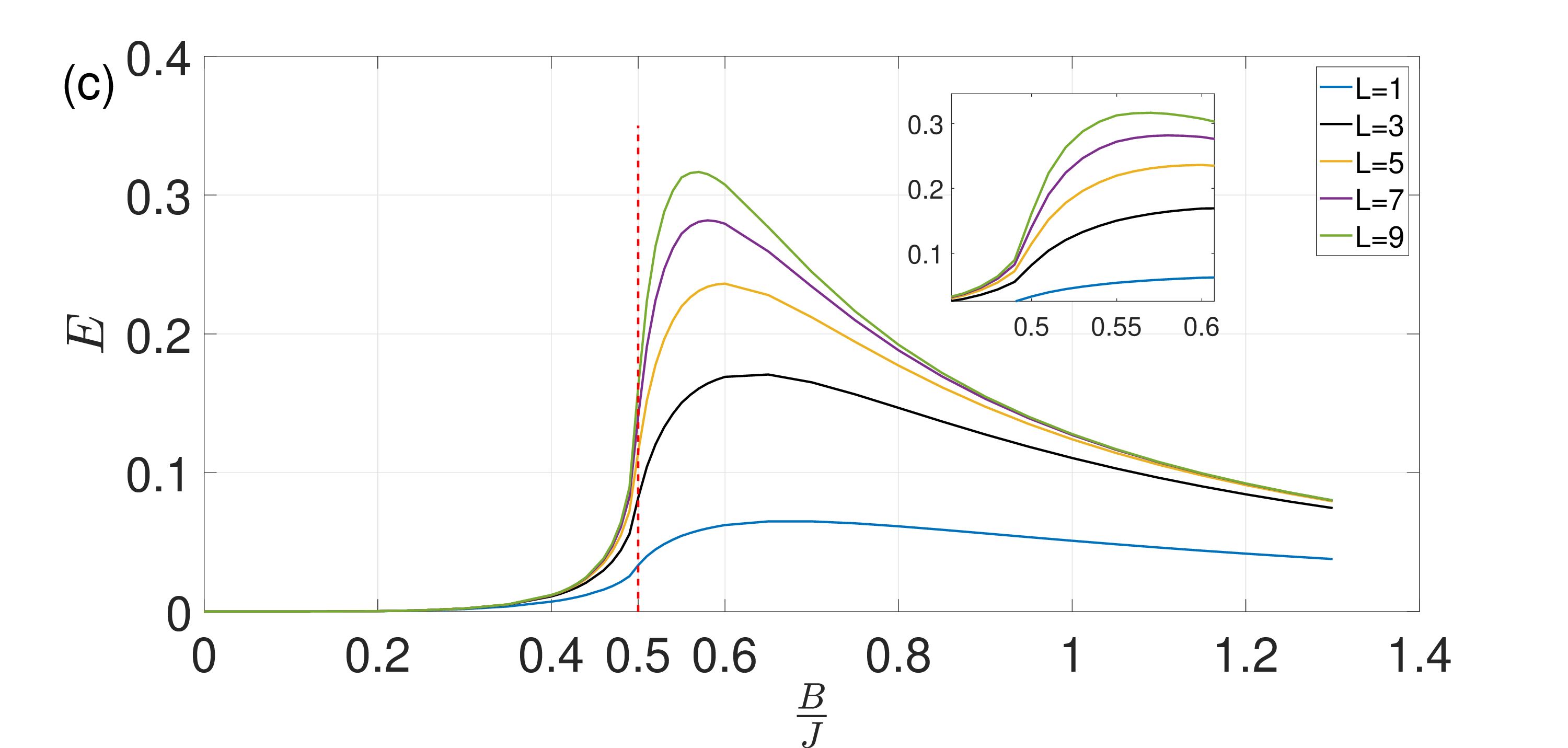}\\
	\includegraphics[width=0.5\textwidth]{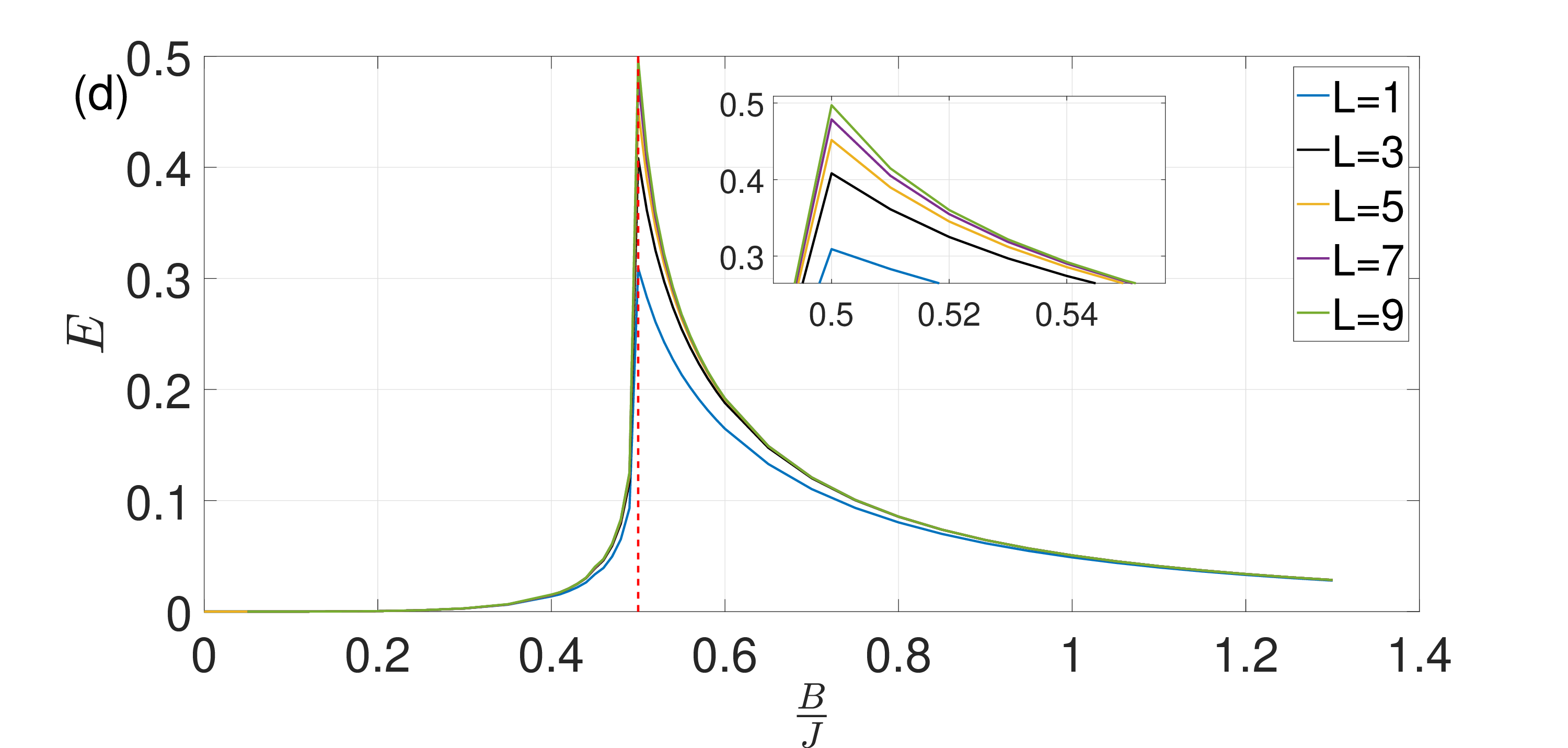}\\
	\caption{$E$ plotted against $B/J$ for sequences of measurements of the quantum Ising chain along (a) $\sigma_{z}$, (b) $\sigma_{x}$, . The vertical dashed lines demarcate the phase transition of the system.}
	\label{fig:E_qi}
\end{figure}

\begin{figure}
	\centering
	\includegraphics[width=0.5\textwidth]{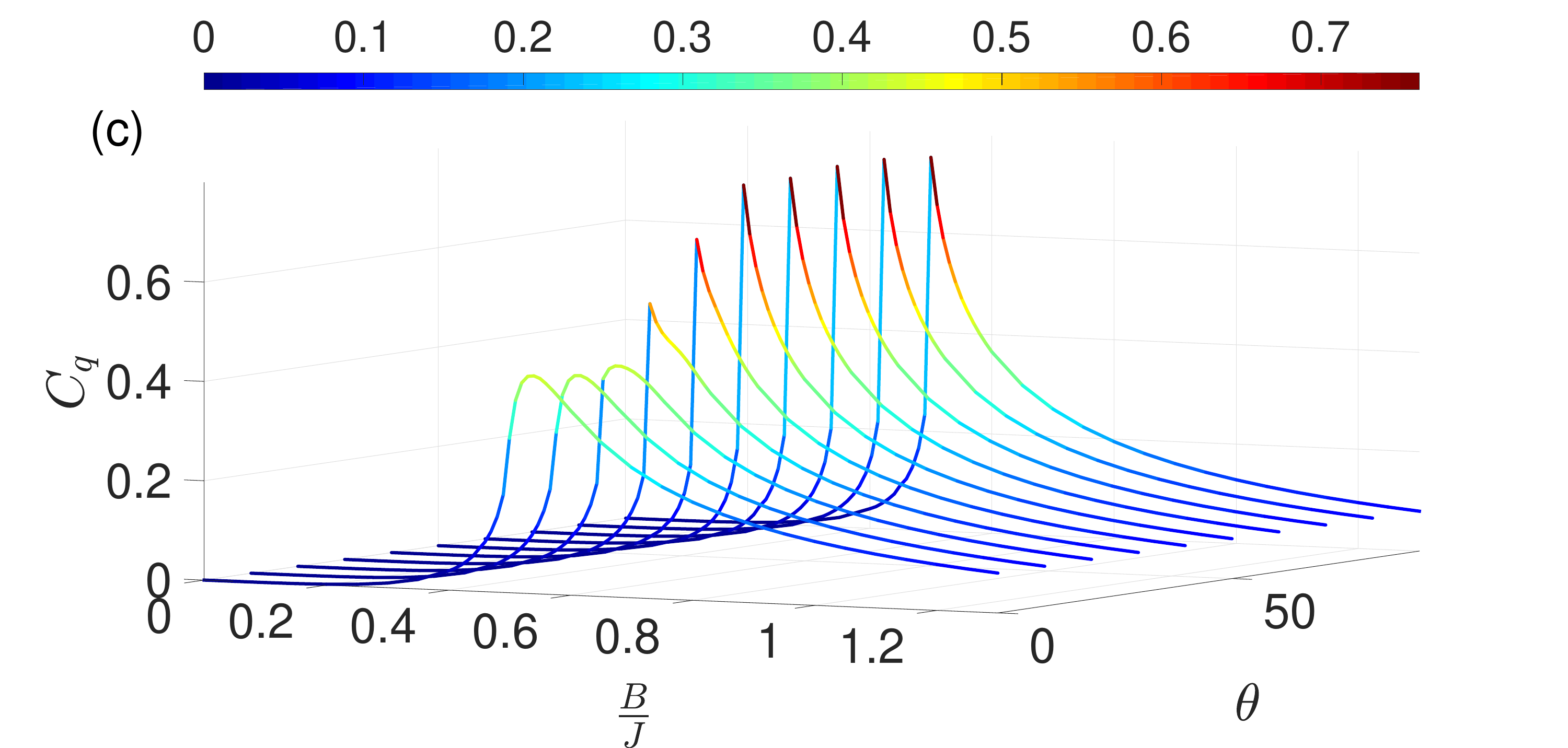}\\
	\includegraphics[width=0.5\textwidth]{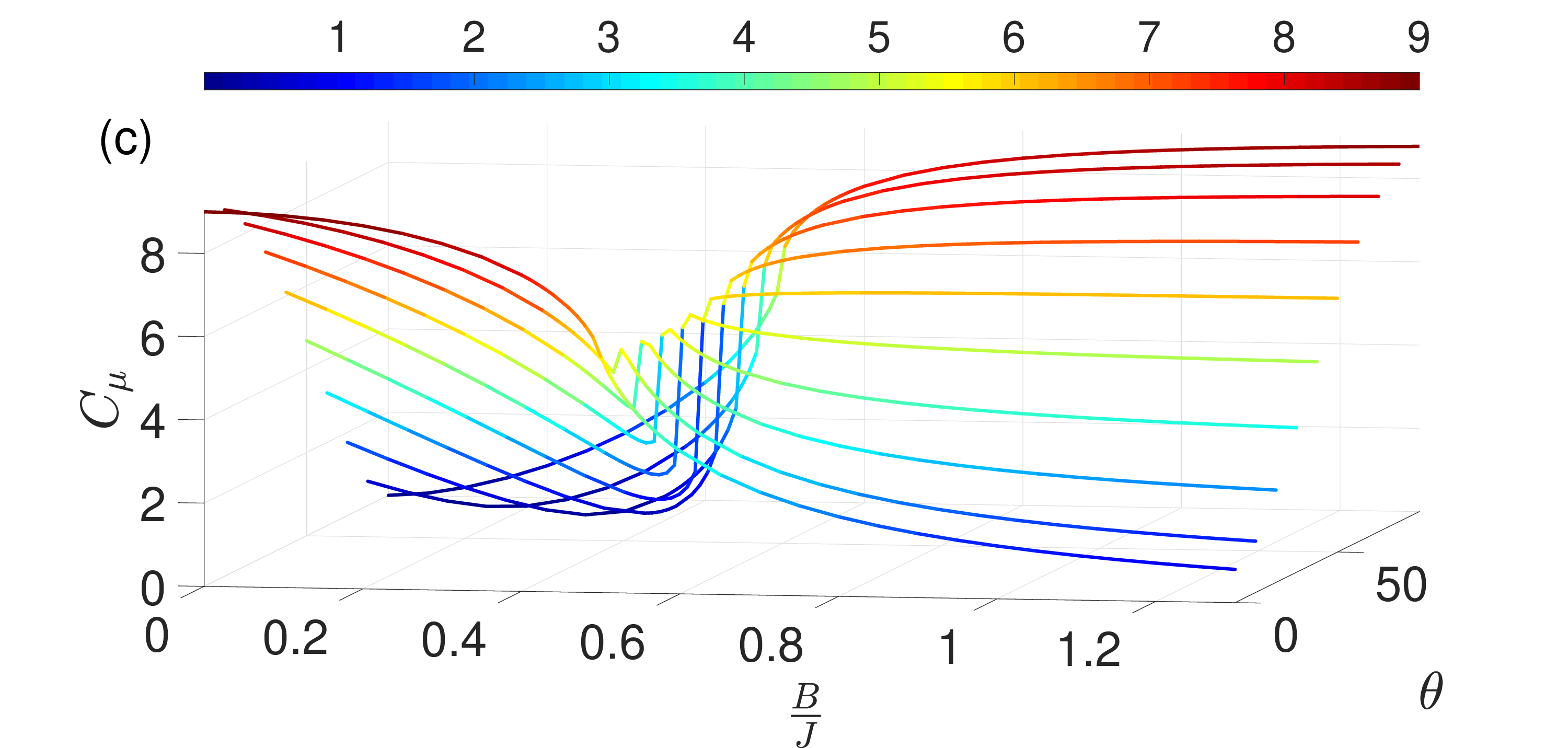}\\
	\includegraphics[width=0.5\textwidth]{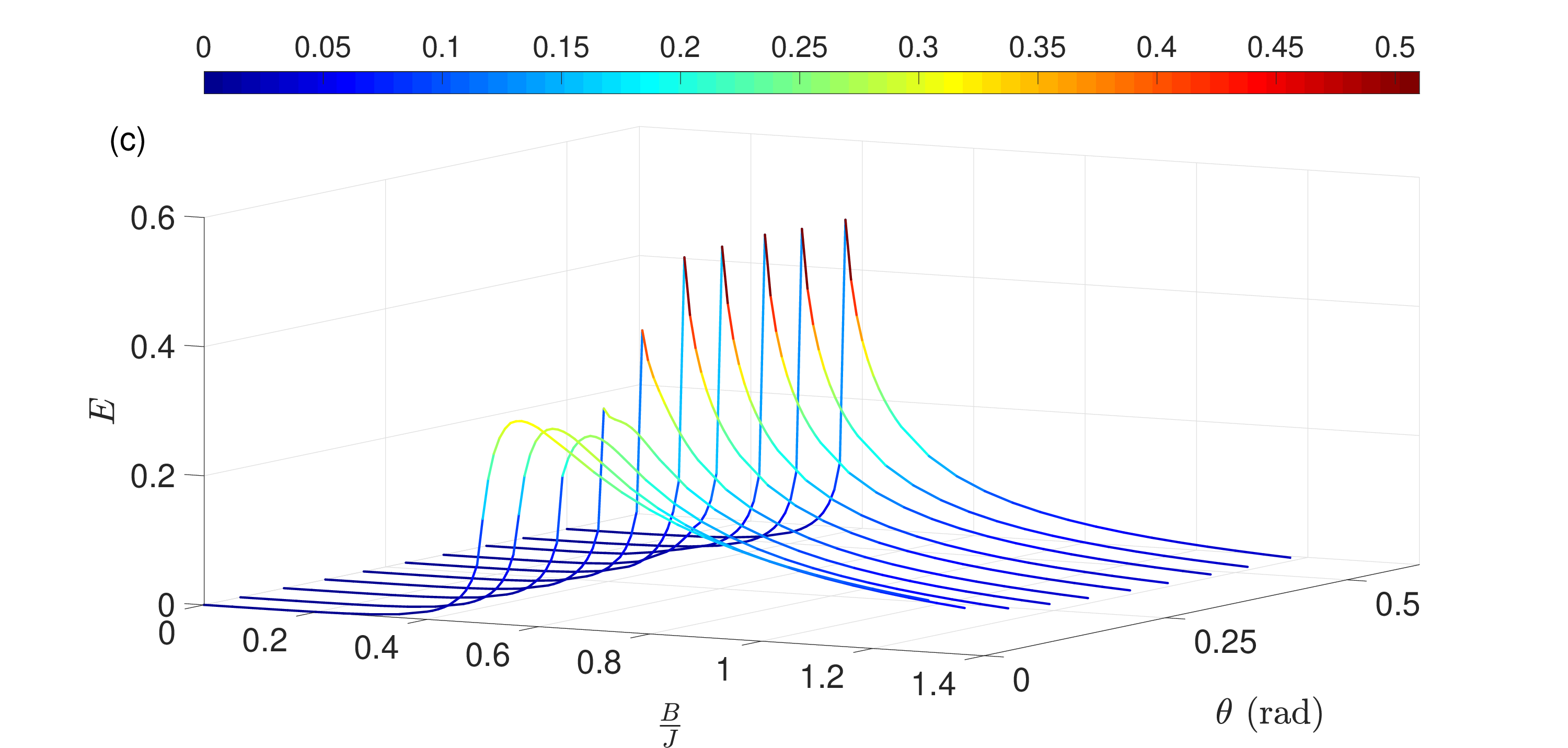}\\
	\caption{Plot of how the behaviours of $C_q$, $C_\mu$, and $E$ changes with respect to $\theta$ (rad).}
	\label{fig:3D_all}
\end{figure}

We also observe differing behaviour of $C_\mu$, $C_q$, and $E$ with respect to the truncated Markov memory order, as illustrated in Figs.~\ref{fig:cq_qi}, \ref{fig:cm_qi}, and \ref{fig:E_qi}. As $L$ is increased, $C_q$ also increases, but at a decreasing rate, indicating that the value of $C_q$ converges as larger $L$ is considered [Fig.~\ref{fig:cq_qi}]. This is consistent with the decrease in correlation strength between spins when the distance between them increases (i.e., spins that are further apart are more weakly correlated). This means that at high $L$, increasing the Markov memory order further adds little predictive information. This cannot be utilised effectively by the classical $\varepsilon$-machine with access only to orthogonal states; as $L$ increases, so too does the number of causal states -- resulting in a higher $C_\mu$, as illustrated by Fig.~\ref{fig:cm_qi}.

We see that the behaviour of $C_\mu$ with respect to $B/J$ changes as we rotate the measurement angle $\theta$ from $\sigma_{z}$ to $\sigma_{x}$ [Fig.~\ref{fig:3D_all}(b)]. As $\theta$ changes from $0$ to $\pi/2$, the measurement outcome sequences sweep between near-random and near-deterministic in the region $B\ll J$, and between near-deterministic and near-random in the region $B\gg J$. Fig.~\ref{fig:3D_all}(a) shows the counterpart behaviour of $C_q$ with respect to the different measurement angles. As the measurement basis changes from $\sigma_{z}$ to $\sigma_{x}$, we observe that the peak of $C_q$ increases with respect to $\sigma_{\theta}$. In the parameter region that is neither random nor deterministic, our quantum model stores less information than when measured in a basis that is closer to the $z$-axis. This captures the underlying structure of the quantum Ising chain very well, as measurement outcomes along the $z$-axis are less dependent on the past spins, because the inter-spin coupling of the system is along the $x$-axis.

\textbf{Bose-Hubbard chain.} We now look at structure in the one-dimensional Bose-Hubbard model, which describes the physics of interacting spinless bosons on a lattice~\cite{jaksch1998cold, greiner2002quantum, lewenstein2012ultracold}. It is governed by:
\begin{equation}
\mathcal{H}_{BH} = -J\sum_l{b^\dagger_l b_{l+1}} + \mathrm{h.c.} +\frac{U}{2}\sum_l{n_l(n_l - 1)}, \label{hamiltonianBH}
\end{equation}
where $b^\dagger_l$ and $b_l$ are bosonic creation and annihilation operators and $n_l=b^\dagger_l b_l$ is the number of bosons at site $l$. The variable $J$ denotes the hopping amplitude,  describing the kinetic energy of the bosons, and $U$ is the on-site repulsive interaction strength. The filling factor $\nu$ is defined as the average number of bosons per site; we consider a chain with $\nu=1$. In our numerical calculations, we use $N=300$ and truncated Markov memory orders of $L\in\{1,2,3,4\}$. We also enforce that each site has a maximum occupation number $n_{\text{max}}=4$ -- higher values of occupation number have very low probability of occurrence (and thus have little impact on the state), yet demand significantly more computational power. In the ground state, when $J\gg U$ all bosons occupy zero-momentum eigenstates, wherein they are completely delocalised across the lattice; this is the superfluid phase. On the other hand, when $J\ll U$, the system is in the Mott insulator phase, where at integer filling factors $\nu$, each site contains $\nu$ highly-localised bosons. The one-dimensional Bose-Hubbard model undergoes a quantum phase transition at $U/J\approx 3.1$~\cite{greiner2002quantum, pino2012reentrance} for the case of $\nu=1$.

\begin{figure}
	\centering
	\includegraphics[width=0.5\textwidth]{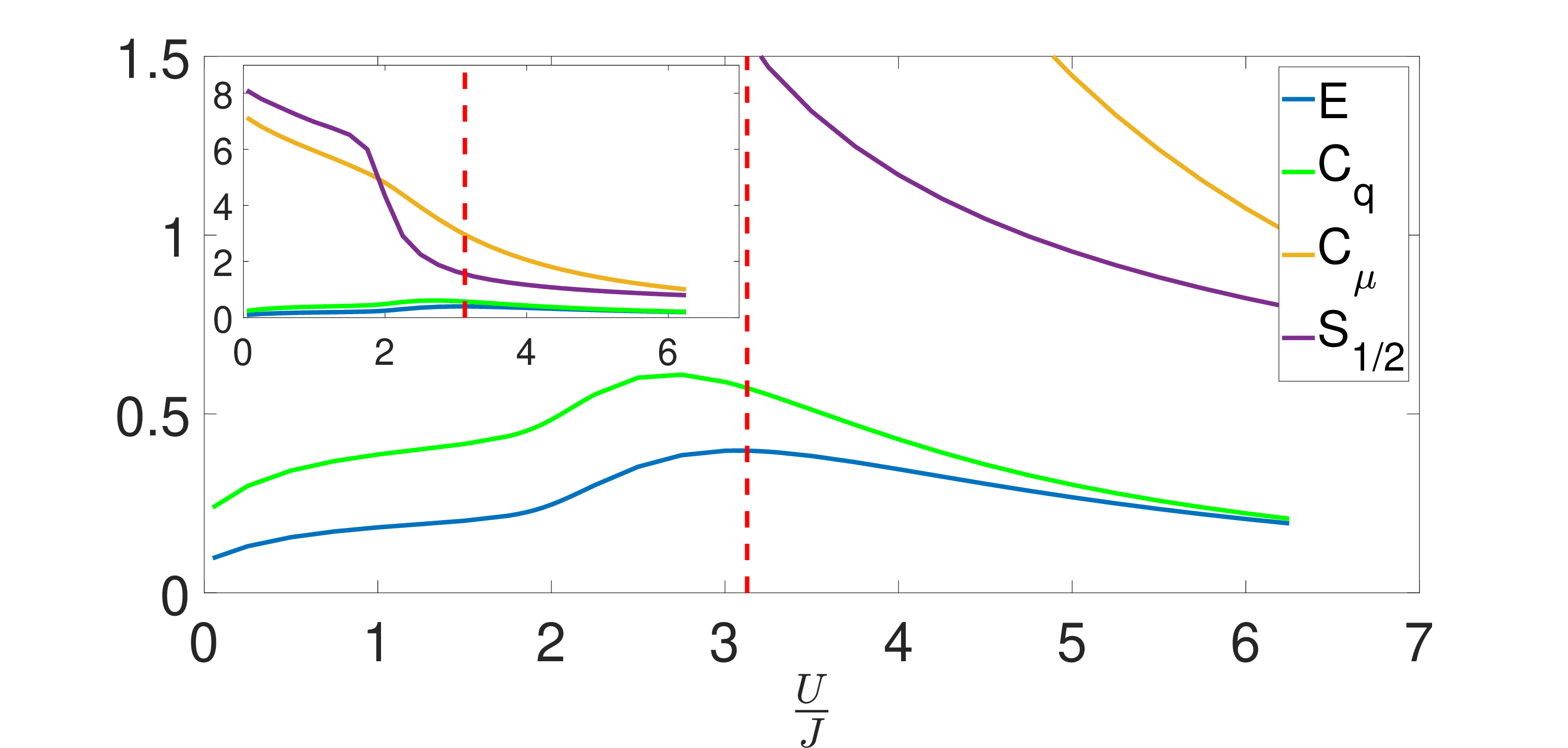}\\
	\caption{$E$, $C_q$ and $C_\mu$ at $L=4$,  and $S_{\frac{1}{2}}$, plotted against $U/J$, for sequences of measurements of $n$ in the Bose-Hubbard chain. The vertical dashed line demarcates the phase transition of the system.}
	\label{fig:IBH}
\end{figure}

\begin{figure}
	\centering
		\includegraphics[width=0.5\textwidth]{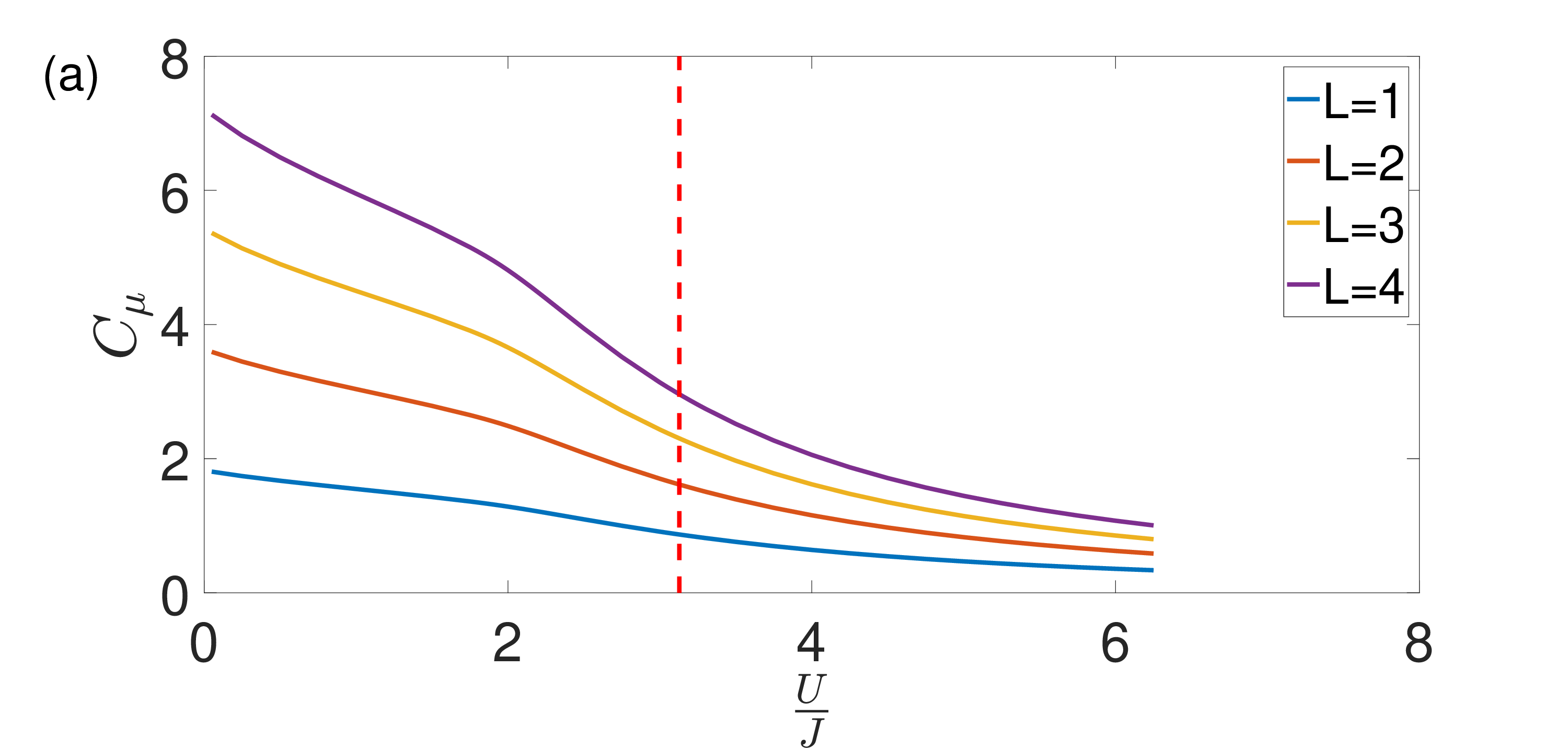}\\
		\includegraphics[width=0.5\textwidth]{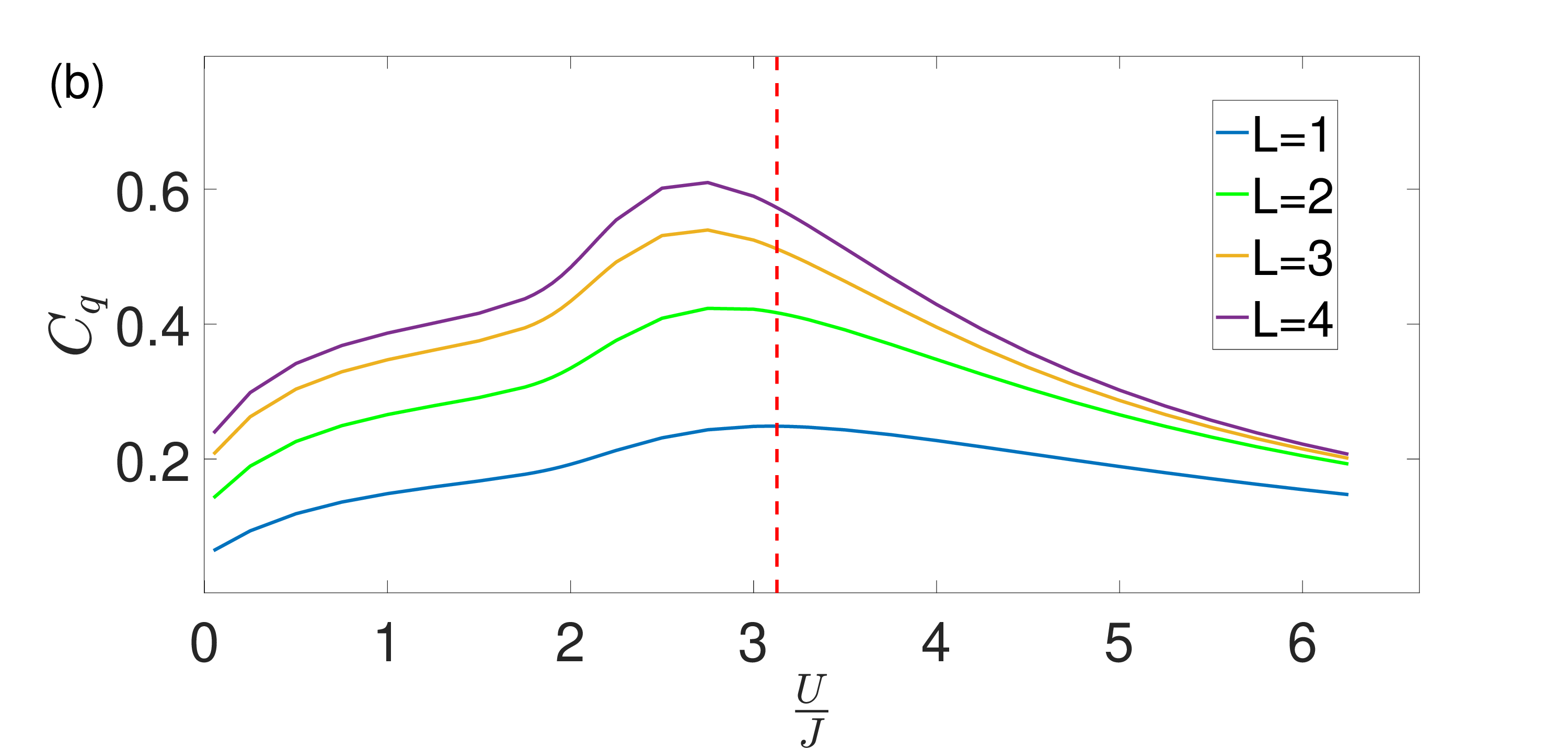}
	\caption{Scaling with $L$ of (a) $C_\mu$ and (b) $C_q$ against $U/J$ for sequences of measurements of $n$ in the Bose-Hubbard chain. The vertical dashed lines demarcate the phase transition of the system.}
	\label{fig:cm_bh}
\end{figure}

We calculate the structural complexity measures for measurement outcome sequences of the Bose-Hubbard chain in the number basis, wherein $n$ is measured sequentially on every site. Fig.~\ref{fig:IBH} shows the  behaviour of $C_\mu$, $C_q$, and $E$ for these sequences, as well as $S_{\frac{1}{2}}$. We observe that $E$ peaks close to the phase transition, while $C_q$ peaks earlier. This indicates that $E$ may identify the phase transition, while $C_q$ and $C_\mu$ do not. This is in contrast to the quantum Ising chain, where both $E$ and $C_q$ (in multiple measurement bases) reach a maximum value in the vicinity of the phase transition. 

In the superfluid phase $(J\gg U)$ $C_\mu$ behaves very differently to $C_q$ and $E$: $C_\mu$ increases as $U/J \rightarrow 0$, while $C_q$ and $E$ decrease. This is because the bosons are delocalised, leading to near-random measurement sequences for the site occupations. Hence, the models occupy all available (quantum) causal states with almost equal probabilities. Analogous to the quantum Ising chain however, $C_\mu$ and $C_q$ behave very differently, due to the (non-)orthogonality of (quantum) memory states. 

On the other hand, in the Mott insulator phase $(J\ll U)$, $C_\mu$, $C_q$, and $E$ decrease as $U/J$ increases. In this regime, the bosons in the system are highly localised, and so measurement in the $n$-basis yields a highly-ordered stochastic process since the number of bosons at each site tends to $\nu$ as $U/J \rightarrow \infty$. Similar to the analogous limit in the quantum Ising chain, measurement sequences that are highly-ordered have a single causal state that manifests with very high probability, while other causal states all occur with low probabilities. This results in low values of both $C_\mu$ and $C_q$.

Fig.~\ref{fig:IBH} also shows that $S_{\frac{1}{2}}$ is larger than $C_q$ for the measurement outcome sequences; as with the quantum Ising chain, this is because $S_{\frac{1}{2}}$ quantifies the full quantum correlations between two halves of the system, while $C_q$ results from projecting the quantum state onto one specific basis, and discarding information about all other bases. Notably, in the superfluid parameter region $S_{\frac{1}{2}}$ behaves very differently to $C_q$ and $E$: $S_{\frac{1}{2}}$ increases, while $C_q$ and $E$ decrease. This is because projecting the state into the $n$-basis removes the structure in the conjugate basis (i.e., momentum), that would have manifest large half-chain entanglement. 

Finally, Fig.~\ref{fig:cm_bh} shows how $C_\mu$ and $C_q$ scale with increasing $L$. As with the quantum Ising chain, we see that $C_q$ displays signs of convergence as $L$ is increased, while $C_\mu$ does not. Again, this is due to the decay in the strength of correlations between site occupations with increasing distance.

\section{Discussion}\label{sec:discussion}
Our results show that classical and quantum measures of structural complexity can exhibit drastically different qualitative behaviour when applied to sequences generated by measurement outcomes of quantum systems. In particular, it is evident that the measures interpret near-randomness very differently; quantum models are typically able to capture the predictive features in near-random sequences without storing a large amount of information about the past, in contrast to corresponding minimal classical models. The quantum measures also appear to signal proximity to a phase transition. For a given Hamiltonian, the most informative (i.e., highest complexity) basis appears to vary with the Hamiltonian parameters; by considering maximisations of $C_q$ over all measurement bases we may obtain a stronger indicator of phase transitions and other related phenomena -- we leave this question for future work. Another interesting open question in this direction is the study of universality classes of quantum systems -- can quantum systems be categorised into universality classes according to their structural complexities? Members of the same universality class have identical critical behaviour despite possibly having radically diverse microscopic behaviour. 

Moving forward in the field of computational mechanics, the transducer framework~\cite{barnett2015computational, thompson2017using, elliott2021quantum} provides a natural extension to study quantum systems as input-output processes, and thus lines up as a natural next step to this work. In such processes, the choice of the measurement basis would form the input, and the resulting measurement sequence is the output, allowing for the fully-quantum nature of non-commuting measurements to be considered. It would be interesting to make this extension and study the dynamics of quantum processes by measuring them at different bases at different timesteps, capturing the structure and complexity within an evolving quantum system. An ultimate goal of quantum computational mechanics is to construct quantum causal models that can simulate any quantum stochastic processes without restriction in choice of measurement bases; the results in this manuscript serve as a crucial first step towards this direction.

\inlineheading{Acknowledgements.}
We thank Felix Binder, Yang Chengran, Jirawat Tangpanitanon, and Benjamin Yadin for enlightening discussion, and the University of Oxford Advanced Research Computing department for providing us with access to their platform to run our numerical simulations. We are also grateful to Sarah Al-Assam, Stephen Clark, and Dieter Jaksch for their permission to use their tensor network library~\cite{al2017tensor}. This work was funded by the Singapore Quantum Engineering Program QEP-SF3, the Singapore National Research Foundation Fellowship NRF-NRFF2016-02, grant FQXi-RFP-1809 from the Foundational Questions Institute and Fetzer Franklin Fund (a donor advised fund of Silicon Valley Community Foundation), the Singapore Ministry of Education Tier 1 grant RG162/19, and the Lee Kuan Yew Endowment Fund (Postdoctoral Fellowship). M.G thanks the FQXi-funded workshop 'Workshop on Agency at the Interface of Quantum and Complexity Science' for catalyzing the research. T.J.E.~thanks the Centre for Quantum Technologies for their hospitality.

\section*{TECHNICAL APPENDIX}
Tensor Network Theory (TNT)~\cite{al2017tensor} is a set of powerful and efficient numerical methods for classically simulating quantum many-body systems. In this Appendix, we briefly review matrix product states (MPS), matrix product operators (MPO), and the density matrix renormalisation group (DMRG) in the context of our work. MPS and MPO provide efficient descriptions of states and operators of quantum many-body systems respectively, while DMRG is an iterative procedure that variationally minimises the energy of Hamiltonians to obtain the ground states of quantum many-body systems.

MPS~\cite{orus2014practical} are widely used as efficient representations of low energy states of one-dimensional quantum systems. In a quantum many-body chain, each lattice site is represented by a tensor, and the tensors are connected to their neighbours. Consider a quantum many-body chain of size $N$ in a quantum state 
\begin{align}
\ket{\psi} &= \sum_{i_1 i_2 \dots i_N}c_{i_1i_2 \dots i_N} \ket{i_1}\ket{i_2}\cdots\ket{i_N}
\end{align}
where  $\{ \ket{i_j} \}$ are the local orthonormal basis states. We can perform repeated Schmidt decompositions~\cite{nielsen2010quantum} at each site, splitting the tensor $c_{i_1i_2 \dots i_N}$ into local tensors $\Gamma^{[j]}$, and Schmidt coefficients $\lambda^{[j]}$ that quantify the entanglement across the split, which gives us the canonical form of the MPS representation of the state:
\begin{align}
\ket{\psi}\!\!&=\!\!\!\!\!\!\sum_{\{  i \},\{  \alpha \}}\!\!\!\!\!\left(\!  \Gamma^{[1]i_1}_{\alpha_1} \!\lambda^{[1]}_{\alpha_1} \Gamma^{[2]i_2}_{\alpha_1 \alpha_2} \lambda^{[2]}_{\alpha_2}\!\ldots\! \lambda^{[N-1]}_{\alpha_{N-1}}\Gamma^{[N]i_N}_{\alpha_{N-1}}\!\right)\!\ket{i_1}\!\ket{i_2}\!\ldots\!\ket{i_N}\!,
\end{align}
where $\alpha_j$ takes positive integer values up to the rank of $\Gamma^{[j]}$. By contracting the Schmidt coefficient tensors $\lambda^{[j]}$ into the local tensors $\Gamma^{[j]}$, we obtain a more generic form:
\begin{align}
\ket{\psi} &= \sum_{i_1 i_2 \dots i_N} A^{x_1}A^{x_2}\cdots A^{x_N} \ket{i_1}\ket{i_2}\cdots\ket{i_N},
\end{align}
where $A^{x_j}$ is a matrix with the same dimension as the local basis states.

In a similar fashion, a quantum operator can be written in the form of MPO~\cite{pirvu2010matrix}:
\begin{align}
\mathcal{H} &= \sum_{i,k} H^{i_1,k_1}H^{i_2,k_2} \cdots H^{i_N,k_N} \ketbra{i_1 i_2\cdots i_N}{k_1 k_2 \cdots k_N},
\end{align}
where $H^{i_j,k_j}$ is a matrix with the dimension of the local basis state. With quantum states and Hamiltonians represented in MPS and MPO forms respectively, ground states $\ket{\psi_g}$ may then be obtained by minimising $\bra{\psi}\mathcal{H}\ket{\psi}$ across all states using the DMRG algorithm.

The DMRG algorithm~\cite{white1992density, white1993density} is an iterative, variational method that truncates the degrees of freedom of the system, retaining only the most significant features required to accurately describe the physics of a target state. The algorithm achieves remarkable precision in describing one-dimensional quantum many-body systems~\cite{white1993numerical}.

In the DMRG algorithm, the elementary unit is a site, described by the state $d_i$ where $i=1,\dots , D$ is the label of the states accessible to a given site. A block $B(L,v_L)$ consists of $L$ sites, and has total dimension $v_L$; $H_B$ is the Hamiltonian of the block, containing only terms that involve the sites inside the block. Whenever a block is enlarged, a site is added to the block, forming an enlarged block $B^e$ with a Hilbert space dimension that is the product of the Hilbert space of $B(L,v_L)$ and a site, i.e.~$v_L \times D$. An important step in the algorithm is the formation of superblock Hamiltonians, consisting of two enlarged blocks connected to each other. The superblock ground state is calculated using Lanczos~\cite{lanczos1950iteration} or Davidson~\cite{davidson1975iterative} methods. The ground state is then truncated by discarding the least-probable eigenstates.

The algorithm itself consist of two parts: the warm-up cycle, and finite-system algorithm. The warm-up cycle is designed to create a system block of the desired length of at most dimension $\chi$, before the finite-system algorithm is applied to compute the ground state. Starting from a block $B(1,D)$, each step of the warm-up cycle is carried out as follows~\cite{de2008density}:

\begin{enumerate}
\item Start from a left block $B(L,v_L)$, and enlarge the block by adding a single site. %$v_L$ is an arbitrarily small number.
\item Form a superblock by adding a reflected copy of the enlarged block to its right.
\item Obtain the ground state of the superblock, and the $v_{l+1}=\min(v_l D,\chi)$ eigenstates of the reduced density matrix of the left enlarged block with largest eigenvalues.
\item The truncated left enlarged block is used for the next iteration.
\item Renormalise all operators to obtain block \mbox{$B(L+1,v_{L+1})$}.
\end{enumerate}

These steps are repeated until the desired length $L_{\mathrm{max}}$ is reached. Once the infinite-system algorithm reaches the desired length, the system consist of two blocks of $B(L_{\text{max}}/2 - 1,\chi)$ and two free sites. The subsequent step is called the ``sweep procedure", the goal of which is to enhance the convergence of the target state. The sweep procedure consists of enlarging the left block with one site and reducing the right block correspondingly to keep the length fixed. While the left block is constructed by the usual enlarging steps, the right block is recalled from memory, as it has been built in the infinite-system algorithm and saved. This procedure is repeated until the left block reaches the length $L_{\text{max}}-4$. At this point the right block $B(1,D)$ with one site is constructed from scratch and the left block $B(L_{\text{max}},\chi)$ is obtained through renormalisation. The sweep procedure is then repeated from right to left, and at each iteration, the renormalised block has to be stored in memory. The procedure is stopped when the system energy converges.

In this manuscript, we use the implementations of these algorithms as described in~\cite{al2017tensor}, and the ground states are computed with $\chi=150$ for the one-dimensional quantum Ising chain, and $\chi=80$ for the one-dimensional Bose-Hubbard chain. The resulting ground states are accurate up to $\mathcal{O}(10^{-14})$.

%\bibliography{reference}

\end{document}